\documentclass[pra,letterpaper,preprint,showpacs,superscriptaddress,floatfix]{revtex4}
\usepackage{graphicx,psfrag,amsmath,amssymb,amsfonts,bbm,latexsym,color,dcolumn,bm}

\begin{document}

\title{Development of an apparatus for cooling $^{6}$Li-$^{87}$Rb Fermi-Bose mixtures\\ 
in a light-assisted magnetic trap}

\author{Michael Brown-Hayes}

\affiliation{Department of Physics and Astronomy,Dartmouth 
College,6127 Wilder Laboratory,Hanover,NH 03755,USA}

\author{Qun Wei}

\affiliation{Department of Physics and Astronomy,Dartmouth
  College,6127 Wilder Laboratory,Hanover,NH 03755,USA}

\author{Woo-Joong Kim}

\affiliation{Department of Physics and Astronomy,Dartmouth
  College,6127 Wilder Laboratory,Hanover,NH 03755,USA}

\author{Roberto Onofrio}

\affiliation{Department of Physics and Astronomy,Dartmouth 
College,6127 Wilder Laboratory,Hanover,NH 03755,USA}

\affiliation{Dipartimento di Fisica ``G. Galilei'',Universit\`a 
di Padova,Via Marzolo 8,Padova 35131, Italy}

\affiliation{Center for Statistical Mechanics and Complexity,INFM-CNR,Unit\`a di Roma 1,Roma 00185,Italy}

\date{\today}

\begin{abstract}

We describe an experimental setup designed to produce ultracold
trapped gas clouds of fermionic $^{6}$Li and bosonic $^{87}$Rb.  
This combination of alkali-metals has the potential to reach deeper 
Fermi degeneracy with respect to other mixtures as it allows for improved 
heat capacity matching which optimizes sympathetic cooling efficiency.  
Atomic beams of the two species are independently produced and then 
decelerated by Zeeman slowers. The slowed atoms are collected 
into a magneto-optical trap, and then transferred into a quadrupole 
magnetic trap. An ultracold Fermi gas with temperature in the 
$10^{-3} \mathrm{T_F}$ range should be attainable through selective 
confinement of the two species via a properly detuned laser beam 
focused in the center of the magnetic trap.
\end{abstract}


\maketitle

\section{Introduction}

The study of superfluidity in ultracold gases has become a rich subfield at 
the border between atomic physics and condensed matter physics \cite{Reviewsbooks}. 
Precision techniques characteristic of atomic physics, controllable environments 
for the dynamics of cold atoms, and the availability of continuous tuning of their 
interactions, have all provided strong impetus for this research which maps into 
the fundamental features of high-temperature superconductivity, still an open problem 
in physics after more than two decades since its discovery \cite{Levin}.  
Superfluidity and vortices in Bose gases have been studied since 1999 
\cite{Matthews,Raman}, and while degenerate Fermi gases were first produced in the same 
year \cite{Jin}, only more recently has Fermi superfluid behaviour been evidenced 
through the generation of vortices in fermionic $^6$Li \cite{Zwierlein1}. 
This time gap in the study of weakly interacting Fermi gases with respect to their 
bosonic counterparts is mainly due to difficulty in the adaptation of successful boson cooling techniques to fermionic species.
In particular, the Pauli principle inhibits efficient evaporative cooling among 
identical fermions when they reach degeneracy.  Two solutions to this basic 
limitation have been realized, namely mutual evaporative cooling of fermions 
in two different states, and sympathetic cooling of the fermions with a Bose 
species having a large heat capacity.  In spite of these ingenious techniques, the smallest 
Fermi degeneracy available to date is in the $T/T_{\mathrm F }$  $\simeq 5 \times 10^{-2}$ 
range \cite{Hadzibabic,Grimm}. This limitation has not precluded the study of interesting 
temperature-independent features of Fermi gases, such as quantum phase transitions related 
to unbalanced spin populations \cite{Zwierlein2,Partridge,Zwierlein3,Shin} and the decoherence 
of a Bose gas in optical lattices filled with Fermi impurities \cite{Gunther,Ospelkaus}.  
However, the study of phase transitions in which temperature is the key parameter is still 
unexplored territory. This will require, as discussed in \cite{Chen,Chien}, the achievement 
of temperatures corresponding to $T/T_{\mathrm F }$ in the $10^{-3}$ range or even lower. 
Unconventional pairing mechanisms which can be unstable at higher temperatures could then be 
observed, and the phase diagram of Fermi atoms in the degenerate regime could be established 
in a wider range of parameter space. Here we describe the strategy and the status of an ongoing 
effort at Dartmouth to reach lower degeneracy factors using fermionic $^{6}$Li 
sympathetically cooled with $^{87}$Rb.  The paper is organized as follows: 
in Section II we briefly recall the current limitations in reaching a deeper 
Fermi degenerate regime and the motivations for choosing the $^{6}$Li-$^{87}$Rb mixture. 
Section III is devoted to a detailed description of the apparatus, including the oven and the 
Zeeman slower for each species, the vacuum chamber, the pumping system, the lasers and 
the frequency synthesis scheme.  In Section IV we describe plans for a light-assisted 
magnetic trap which should allow us to reach the targeted Fermi degeneracy. 
In the conclusions, we place our research project in the broader context of the study of 
ultracold Fermi gases, and  we briefly discuss future physics goals.

\section{Limitations to reach deeper Fermi degeneracy}
As mentioned earlier, all experiments studying ultracold Fermi gases have so far been 
unable to reach a temperature lower than about 5 $\%$ of the Fermi temperature, $T/T_{\mathrm F }$=$5 \times 10^{-2}$, 
and it is important to understand the limiting factors.  In the case of dual evaporative cooling, 
one performs a selective removal of the most energetic fermions in both the hyperfine states. 
Provided that the initial number of atoms in each state is roughly the same, efficient 
dual evaporative cooling can be performed since the heat capacities are comparable during the 
entire process. 
This has the drawback that the number of available atoms progressively decreases over time, 
and correspondingly so does the Fermi temperature (proportional to $N_{F}^{1/3}$).  
As a result, the relative gain in terms of a lower $T/T_{\mathrm F }$ ratio will be 
marginal, and the smaller clouds obtained at the end of the evaporative cooling can 
be detrimental to detailed experimental investigations requiring a large number of atoms.  
In the case of Bose-driven sympathetic cooling, the number of fermions is instead kept 
constant (apart from unavoidable losses due to background pressure and two and three-body 
collisions), and the cooling efficiency now depends on the comparison between the fermion 
heat capacity and that of the Bose coolant.  Unfortunately, in the degenerate regime the 
heat capacity of a Bose gas decreases faster than that of the Fermi gas as the cooling 
proceeds towards the lowest  temperatures. Even in a simplified thermodynamic approach, with 
non-interacting Fermi and Bose gases, the crossover between the heat capacities (assuming 
equal trapping strengths and similar number of atoms for the two species) occurs around 
5-30$\%$ of the Fermi temperature \cite{Presilla,OnofrioJSP}, in line with what has been 
experimentally observed so far.  
One solution to this issue consists in intentionally unbalance the degeneracies, 
keeping the Bose gas less degenerate ({\it i.e.} more classical) with respect to the 
Fermi gas.  More classicality can be achieved by using a more massive species and/or 
by weakening the harmonic confinement of the Bose species relative to that of its Fermi 
counterpart \cite{Onofrio}.  Such species-selective trapping may be achieved via  a light 
beam that is blue-detuned only relative to the Bose species. To quantitatively assess the 
cooling strategy we have examined the choice of species and determined optimal  trapping parameters \cite{mbh1}.  
Our comparison was narrowed to fermionic $^6$Li cooled through bosonic $^{23}$Na, $^{87}$Rb, and $^{133}$Cs, but 
the results can be generalized to other combinations of species such as those involving $^{40}$K as the 
fermionic counterpart and also species recently brought to degeneracy, $^{3}$He \cite{McNamara} and 
$^{173}$Yb \cite{Fukahara}. A significant improvement in the cooling efficiency can be achieved 
for large trapping frequency ratios, with gravitational sagging providing an upper bound, such 
that there will be an optimal value of the trapping frequency ratio in the range of 
$\omega_{\mathrm F}/\omega_{\mathrm B} \simeq 10$. It turns out that the $^6$Li-$^{87}$Rb 
mixture outperforms the other two possibilities, $^6$Li-$^{23}$Na and $^6$Li-$^{133}$Cs, as 
it has a large mass ratio with respect to the former, and at the same time mitigates the 
gravitational sagging with respect to the latter.  Additionally, the cooling efficiency 
benefits from the larger spatial overlap between the two species, and from the mitigation 
of Fermi-hole losses \cite{Eddy}. 

\section{An Apparatus for Trapping and Cooling $^{6}$Li and $^{87}$Rb}

\begin{figure}[t]
\includegraphics[width=1.00\columnwidth]{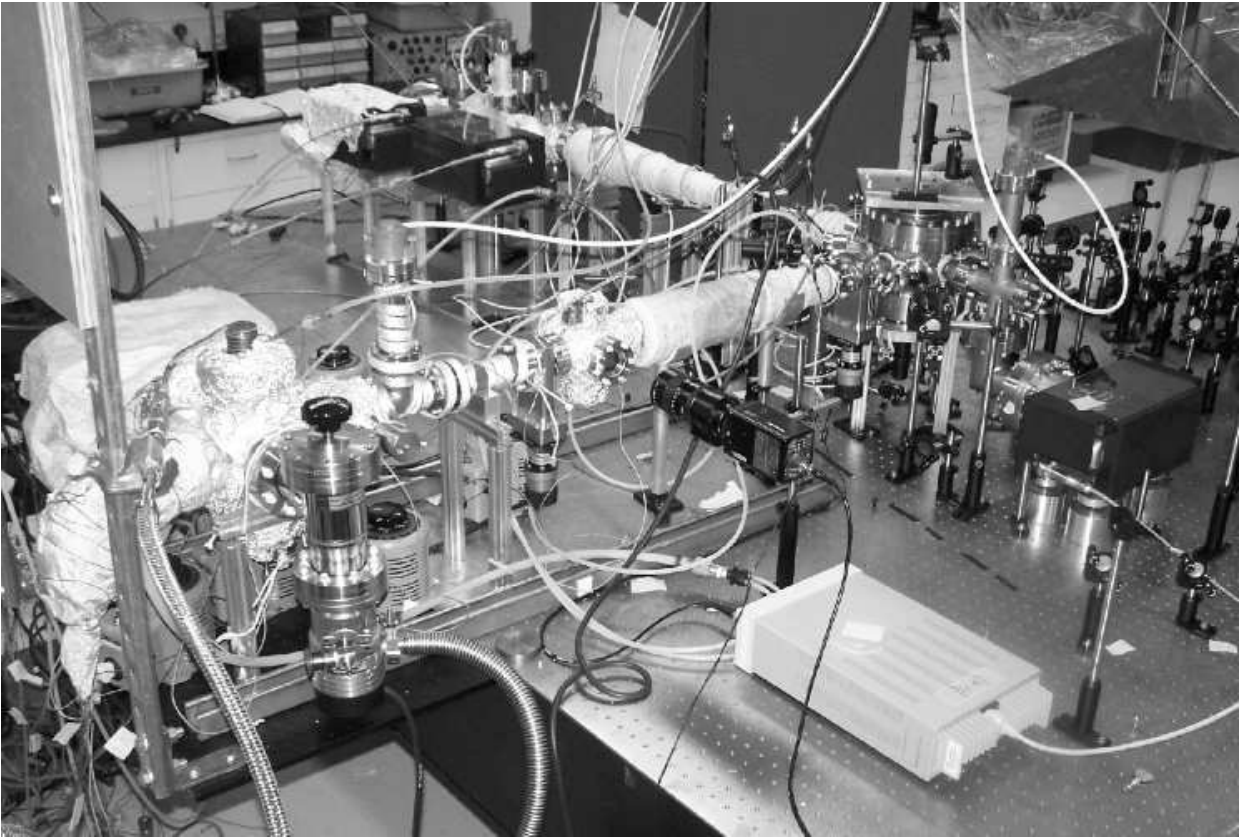}
\caption{Picture of the vacuum system, with the two orthogonal beamlines 
for $^{87}$Rb (top left) and $^6$Li (center). The two slowers converge 
into the vacuum chamber for simultaneous trapping of the mixture.  
The apparatus can be operated in single-species mode by switching off gate valves, one 
of which is visible on the lithium line just before the main chamber.} 
\label{vacuum-system}
\end{figure}

\begin{figure}[t]
\includegraphics[width=1.00\columnwidth]{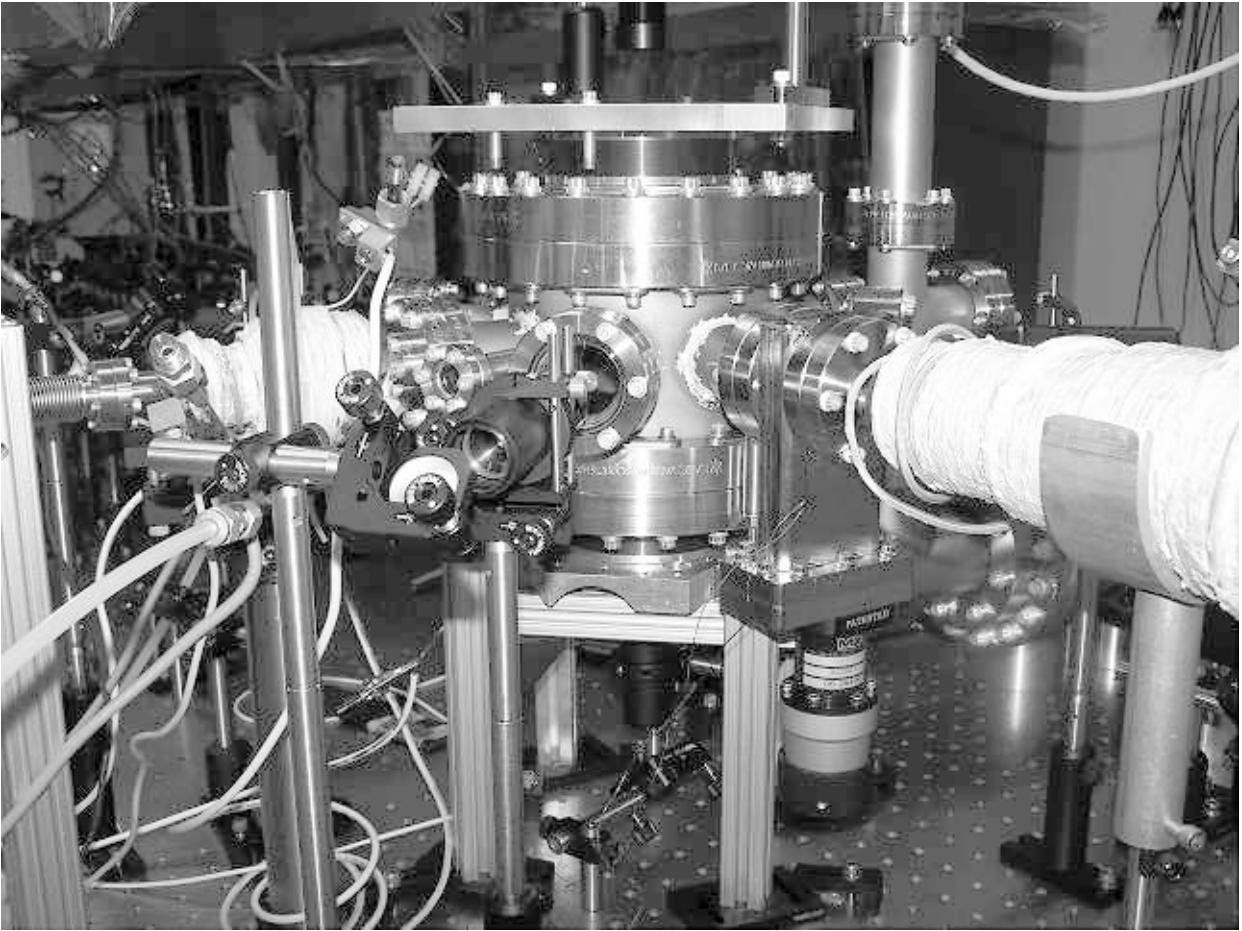}
\caption{Close-up picture of the MOT vacuum chamber. 
On the left is the small coil for the increasing field part of the Rb Zeeman slower, while 
on the right side the final part of the decreasing field Zeeman slower for Li is visible.  
The vacuum chamber is sandwiched above and below by the MOT coils, together with aluminum platforms 
used to mount optics and to provide large surfaces for air cooling of the MOT coils.}
\label{trapping-chamber}
\end{figure}

Our system is constructed with individual effusive ovens and Zeeman slower beamlines 
for $^{6}$Li and $^{87}$Rb which are then joined in a common trapping chamber.  
This configuration has several advantages, including a simpler, single-species oven 
design and greater ability to optimize for each species, as well as allowing 
for independent operation of the beamlines.  Single-species ovens are 
also more practical due to the large difference in operating temperatures for lithium 
(450$^\circ$C) and rubidium (100$^\circ$C), as compared to double-species ovens for 
lithium and sodium operating at 390$^\circ$C and 360$^\circ$C respectively \cite{Stan}.  
In the only other experiment using $^{6}$Li and $^{87}$Rb carried out at T\"ubingen, a Zeeman 
slower for lithium was used in conjunction with a rubidium getter to achieve Fermi and 
Bose degeneracy \cite{Silber}. With respect to the experiment in T\"ubingen, our approach 
should allow for a larger ensemble of rubidium atoms and therefore a greater cooling capability.
Figure \ref{vacuum-system} illustrates the main part of the apparatus.  
For each species, an effusive atomic beam is generated in the oven and Zeeman slowed before 
reaching the common science chamber, where the atoms are trapped in a magneto-optical trap (MOT).   
The main vacuum chamber is shown in Fig. \ref{trapping-chamber} consisting of orthogonal pairs 
of 2-3/4 inch CF flanges for the lithium and rubidium beamlines and Zeeman viewports, as 
well as six flanges for the MOT viewports.  All viewports on the main chamber have a broadband 
anti-reflective coating centered around 700 nm.  The main chamber also has four 1-1/3 inch viewport 
flanges for optical diagnostics.  

\subsection{Effusive Atomic Ovens}

\begin{figure}[t]
\includegraphics[width=1.00\columnwidth]{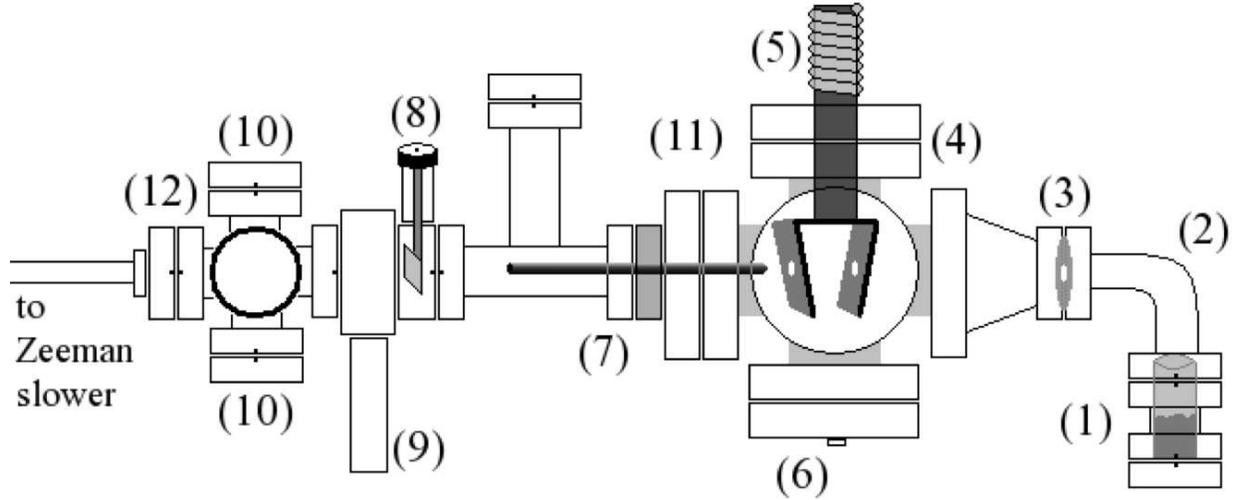}
\caption{Schematic of rubidium oven and beam preparation chamber.  The metal is placed in a sample 
cup (1) which is inserted into the oven nipple (2).  A blank copper gasket (3) with a 3 mm center 
hole provides the first element of collimation.  Further collimation is provided by a double coldplate 
(4) attached to the LN$_{2}$-cooled copper cold finger (5).  Argon/dry nitrogen is pumped in through 
(6) during atom sample changes.  A differential pumping tube (7) allows for the required pressure 
gradient between oven and science chamber, with beam shutter (8) and gate valve (9) providing beam 
and vacuum monitoring.  Pre-slower diagnostic viewports (10) also help with rough alignment, and 
ion pumps are indicated as behind the large (11) and diagnostic (12) 6-way crosses.
}
\label{oven}
\end{figure}

Atoms escape through a 3 mm hole made in a blank copper gasket part of the heated oven, forming 
an effusive atomic beam after passing through collimation elements.  Rubidium has a higher vapor 
pressure ($\sim 3 \times 10^{-7}$ Torr at 100$^{\circ}$C) as compared to lithium or sodium, which is 
incompatible with the UHV  requirement for the science chamber, and as such a careful oven 
design is needed.  A double cold-plate (item 4 in Fig. \ref{oven}) cooled to $\simeq$ 0$^{\circ}$C 
through a liquid nitrogen (LN$_2$) cold finger to reduce atomic vapor in the oven chamber, both improving the 
operating vacuum and reducing the degree of potential alkali-poisoning of the ion pump during operation.  
After baking out the chamber, we have achieved pressures lower than $7 \times 10^{-9}$ Torr.  
Even with the oven running at the extreme temperature of 180 $^{\circ}$C (compared to 
nominal operating temperatures of 110-150 $^{\circ}$C) the cold-plate/ion pump combination 
maintained a pressure of $8 \times10^{-7}$ Torr. 
The estimated differential pumping ratio is $\simeq 10^2$, with the Zeeman slower tube itself 
providing another factor of 4.  We have measured a total pressure ratio of $10^{3}$ between the 
oven and the MOT chamber and believe that the presence of minor leaks in the MOT chamber is the 
current limiting factor. The current vacuum in the main chamber is still sufficient for efficient 
operation of the MOT.

Rubidium is highly reactive and if exposed to air in the standard laboratory environment can 
present a serious safety hazard. As such, an improvised tent is sealed around the oven region and 
kept at positive Ar pressure throughout the loading procedure, which consists of breaking a sealed glass 
5 g rubidium ampule, placing the rubidium in the oven cup, then sealing the oven elbow flange, all within the Ar tent.  
In other experiments \cite{Streed, Schwindt_thesis} the ampule is inserted and then broken under vacuum, 
which has the advantage of a cleaner Rb sample but is mechanically more complicated and has a 
higher leak risk due to the use of a bellow.  In either case it is vital that the exterior 
of the glass ampule is cleaned thoroughly to prevent contamination since even a small amount 
of rubidium oxide will seal the sample and significantly reduce the atomic flux.  
We have also succeeded in melting the rubidium (melting point = 39.31 $^{\circ}$C) while still 
inside the glass ampule and pouring it into the sample cup which is then inserted into 
the elbow, again with the whole procedure conducted inside the positive-pressure Ar tent.

The lithium oven is constructed in the same manner as for rubidium, though with a few additional considerations.  
As already noted, lithium has a lower vapor pressure with respect to rubidium at a given temperature, and this 
requires a much higher operating temperature (450$^\circ$C) to obtain a beam of comparable flux.  
At these temperatures the copper gaskets can bond with the flange knife edges, while lithium can 
alloy with nickel gaskets and can potentially diffuse through them \cite{Stan}.  Careful selection 
of nickel gaskets and the use of 316 stainless steel flanges solve the problem.  In our case we do 
not require high atomic flux - $10^{6} \div 10^{7}$ trapped atoms in the MOT should be sufficient for 
our initial physics goals - and therefore the oven may be operated at lower temperatures.

\subsection{Zeeman slowers}

We use Zeeman slowers for both atomic beams, compensating for the Doppler shift of the progressively 
slowed atoms with a Zeeman shift due to an external magnetic field, such that the atoms are always 
kept in resonance on the cyclic transition.  Zeeman slowers are relatively simple to construct, usually 
more efficient than other methods such as chirped \cite{Ertmer} or broadband \cite{Zhu} slowing, and allow 
for the continuous slowing of ~10$^9$ atoms/sec from (300-1200 m/s) to the MOT capture velocity (10-60 m/s).  
The maximum deceleration of the atoms in the Zeeman slower with infinite laser power is 
a$_{\mathrm{max}}=\hbar k\Gamma/2m$ with natural linewidth $\Gamma$ (6.06 MHz and 5.87 MHz for 
the $^{87}$Rb and $^{6}$Li D2 transitions, respectively) and atomic mass $m$; 
a$_{\mathrm{max}} = 1.085 \times 10^{5}$ m/s$^{2}$ for Rb and $1.82 \times 10^{6}$ m/s$^{2}$ for lithium.  
A decreasing field slower is used for lithium, while for rubidium we use a spin-flip slower.  
The latter combines the advantages of a decreasing field (smaller laser detuning) with increasing field 
(rapid decay of fringing fields) slower designs, at the cost of a slightly higher complexity in the construction.  

\begin{figure}[t]
\includegraphics[width=1.00\columnwidth]{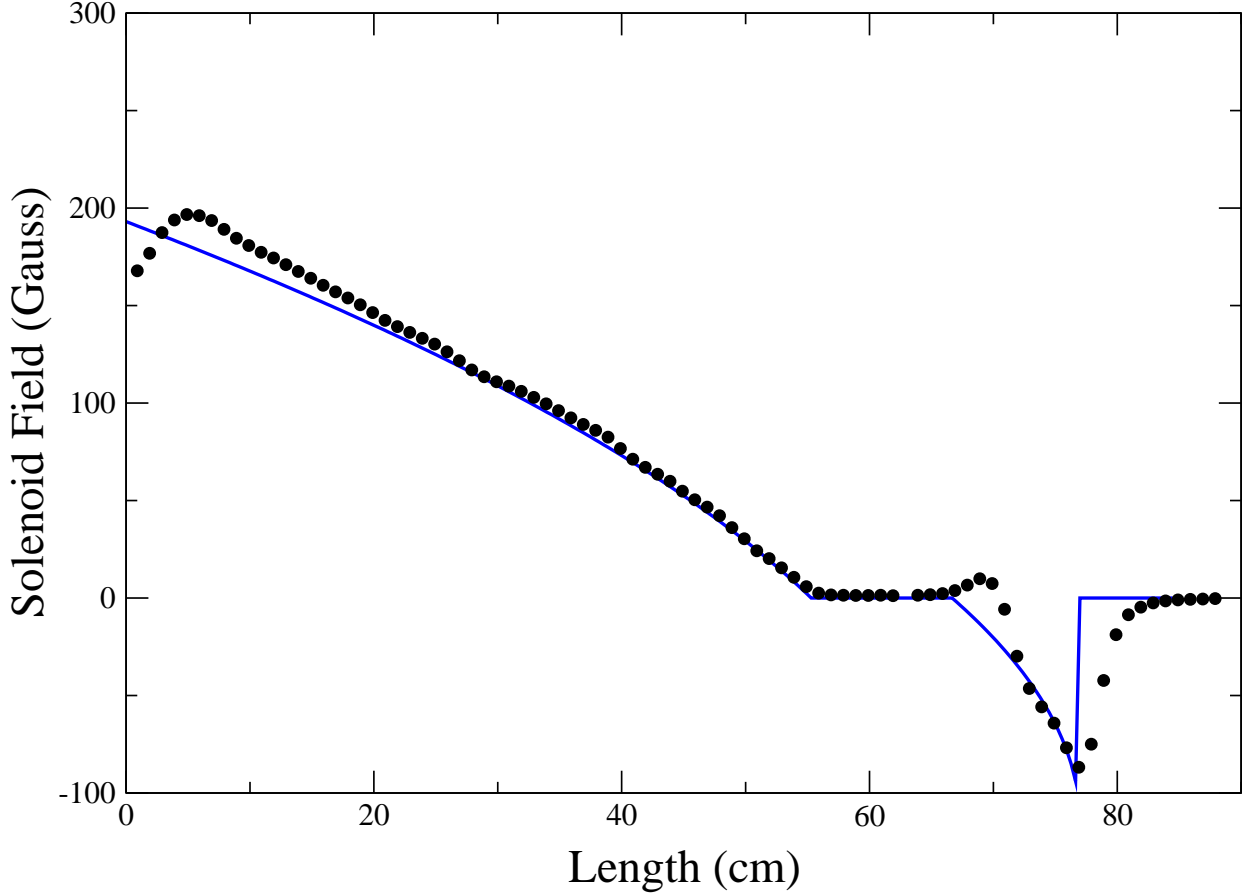}
\caption{Magnetic field profile of the rubidium spin-flip Zeeman slower, showing theoretical (solid curve) and 
measured (dots) profiles.  The theoretical magnetic field versus the distance $x$ along the slower measured from 
its beginning  is given by $B(x)=B_0\sqrt{1-2 {\mathrm a} x/v_{i}^{2}}$ with an offset for the gap between 
the two spin-flip sections, for an initial velocity $v_{i}$=340 m/s, B$_{0}$=310 G, and a laser power 
of $4I_{o}$ which yields a deceleration a=0.8a$_{\mathrm{max}}$.}
\label{rbslowerprofile}
\end{figure}

The optimal length for the rubidium slower was evaluated to be $\simeq$ 70 cm; a more cautious 
design called for 80 cm, allowing for finite laser power and imperfections in the slower solenoid.  
Atoms are slowed from $\simeq$ 340 m/s to roughly 50 m/s.  The main portion of the slower is 57 cm 
long consisting of up to 10 layers of square hollow-core copper wire, wound on a 1 inch OD brass tube, 
building up the desired magnetic field profile.  The second smaller 13 cm section consists of a 
two-layer bias solenoid from which the desired profile is subtracted with a second, increasing 
field solenoid.  Two small countercoils are used to zero the magnetic field in the 
gap between the two sections.  The magnetic field profile is shown in Fig. 
\ref{rbslowerprofile} (measured off-line, before installation) along with the ideal target profile.  
To allow for both ease of installation and oven-slower-chamber alignment corrections, a 10 cm long 
1-1/3 inch CF baffle is used between the two slower sections. 

As we use a decreasing magnetic field slower, the slower light beam is $\sigma^{+}$-polarized, 
utilizing the cycling transition F=2, m$_{F}$=2$\to$F$^\prime$=3, m$_{F}$=3 (see Section \ref{optics} 
for details of the optical pumping scheme).  The longitudinal magnetic field inside the slower 
provides a well-defined quantization axis and the combination of slower beam polarization and 
Zeeman splitting serves to minimize losses due to atoms being optically pumped out of the 
cycling transition.  Another possible concern is transverse heating \cite{guntherwebpage}.  
For the case of atoms slowed by $v_i-v_f \simeq 330$ m/s over 1.1 m, and $v_f$=15 m/s, the 
relative rms velocity is $\Delta v_{x,y}/v_f=5 \%$, a potentially significant loss of collimation, 
which can be mitigated through the implementation of transverse cooling beams sent through  
the 6-way cross.

\begin{figure}[t]
\includegraphics[width=1.00\columnwidth]{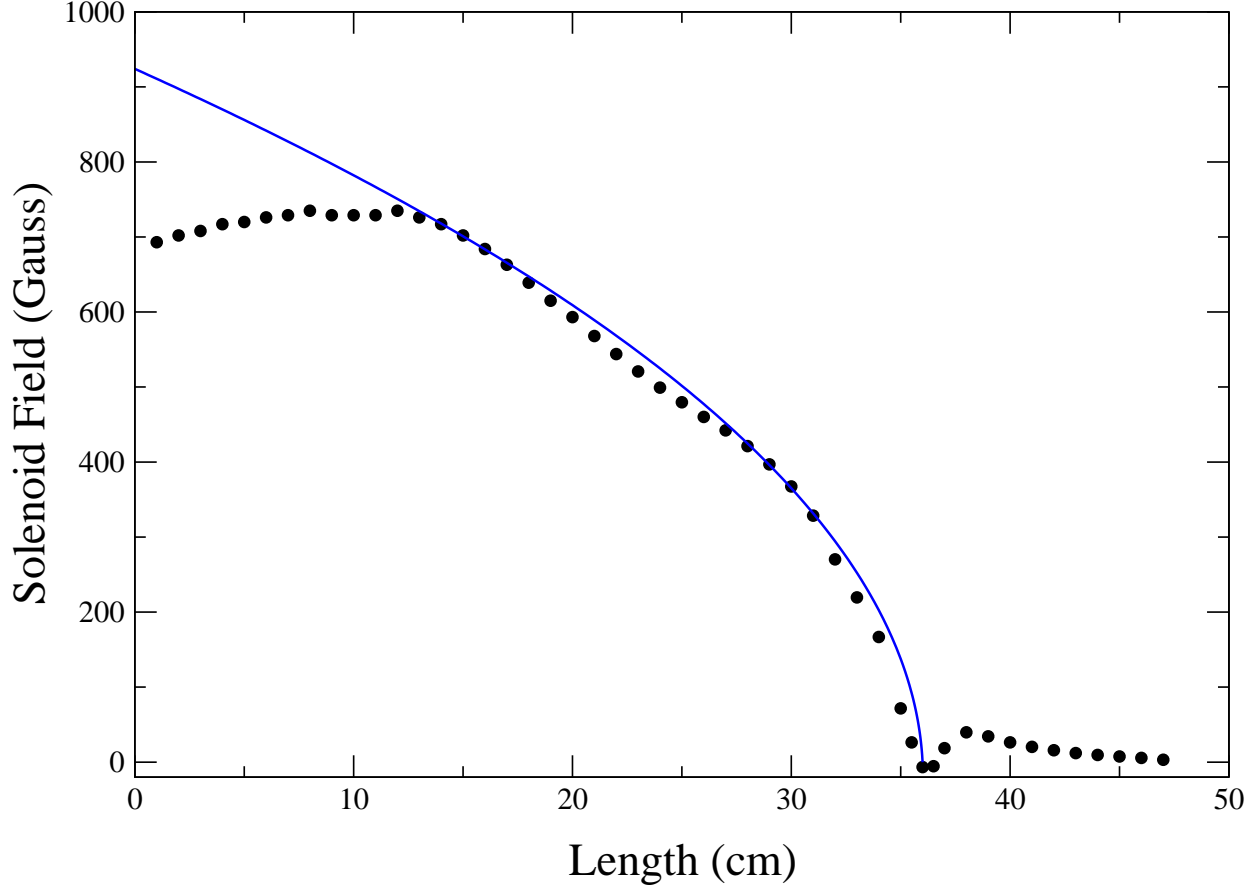}
\caption{Magnetic field profile of the lithium Zeeman slower, showing theoretical (solid line), 
and measured profiles (dots).  The theoretical profile was conservatively calculated 
for atoms with initial velocity of 724 m/s, requiring a laser power of 
$1.5 I_{0}$, a=0.6 a$_{\mathrm{max}}$. The difference between the targeted profile 
and the measured one in the first 10 cm may be minimized by wrapping three extra layers 
of coils, which will increase the maximum capture velocity of the Zeeman slower to 887 m/s.
}
\label{lislowerprofile}
\end{figure}

The much lower mass of lithium, combined with the higher operating temperature of the oven, results 
in a much higher initial velocity of $v_{i} \sim$ 1000 m/s.  This requires a much stronger magnetic field 
(920 G) as compared to rubidium (310 G), suggesting the use of a decreasing-field slower to 
minimize fringing magnetic fields at the end of the deceleration stage.  
Our slower solenoid is constructed similarly to the main section of the rubidium slower, with 
1/8 inch square, hollow copper wire wound around a 1.5 inch brass mounting tube.  
The slower is 36 cm long, starting with 10 layers decreasing down to 2, with a small countercoil at the 
end to limit fringing magnetic field effects.  The comparison between the desired profile with 
the measured profile is shown in Fig. \ref{lislowerprofile}. The lithium slower and main and bias 
sections of the rubidium slower operate at currents of 35, 11, and 30 A, respectively, with the 
countercoils for both slowers running in the 1-5 A range.  These currents are provided by 600 W- 1kW Agilent 
and Kepco power supplies, which require water cooling of the coils.

\begin{figure}[t]
\includegraphics[width=1.00\columnwidth]{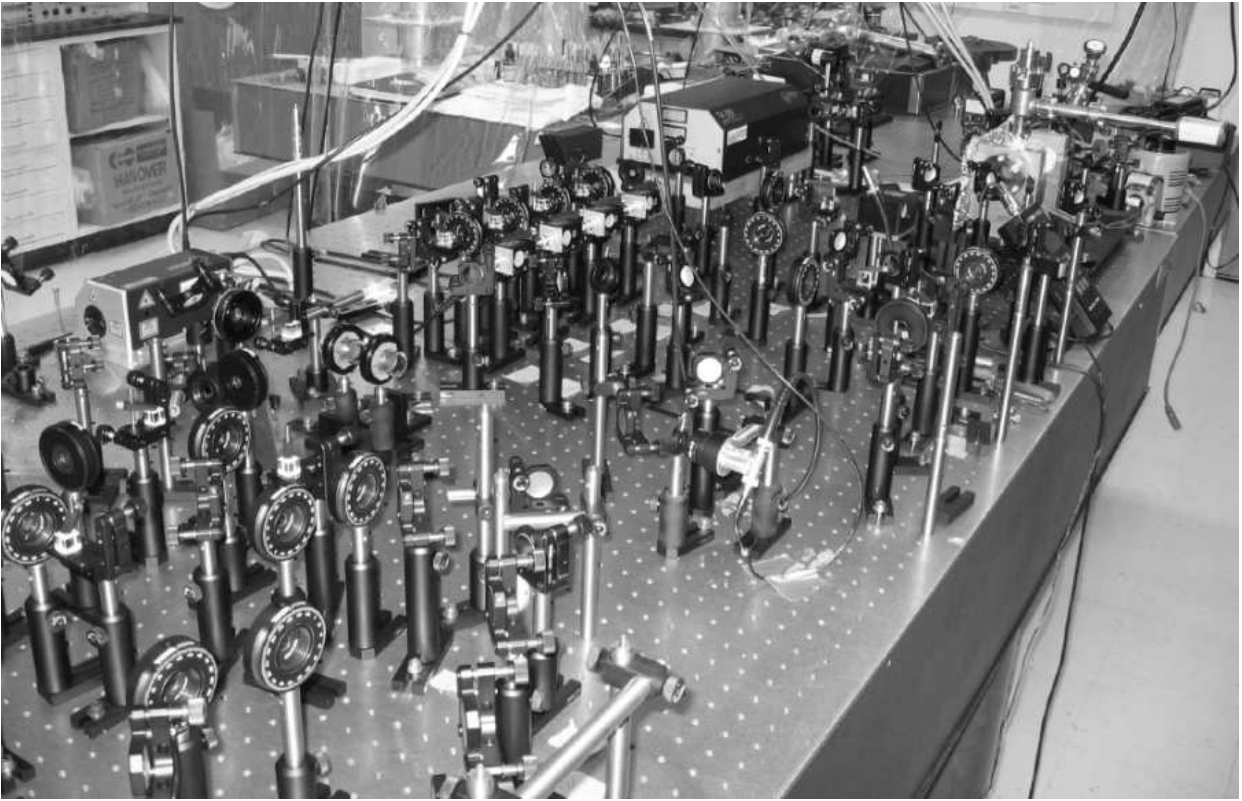}
\caption{Optical table with 671 nm and 780 nm laser systems, Rb and Li vapor  cells 
for saturated-absorption frequency locking, and optical elements for frequency synthesis.  
On the left is the 780 nm laser and related Rb vapor cell, while in the top center is the 
671 nm laser with the related Li vapor cell on its right.}
\label{LaserPhoto}
\end{figure}

\begin{figure}[t]
\includegraphics[width=1.00\columnwidth]{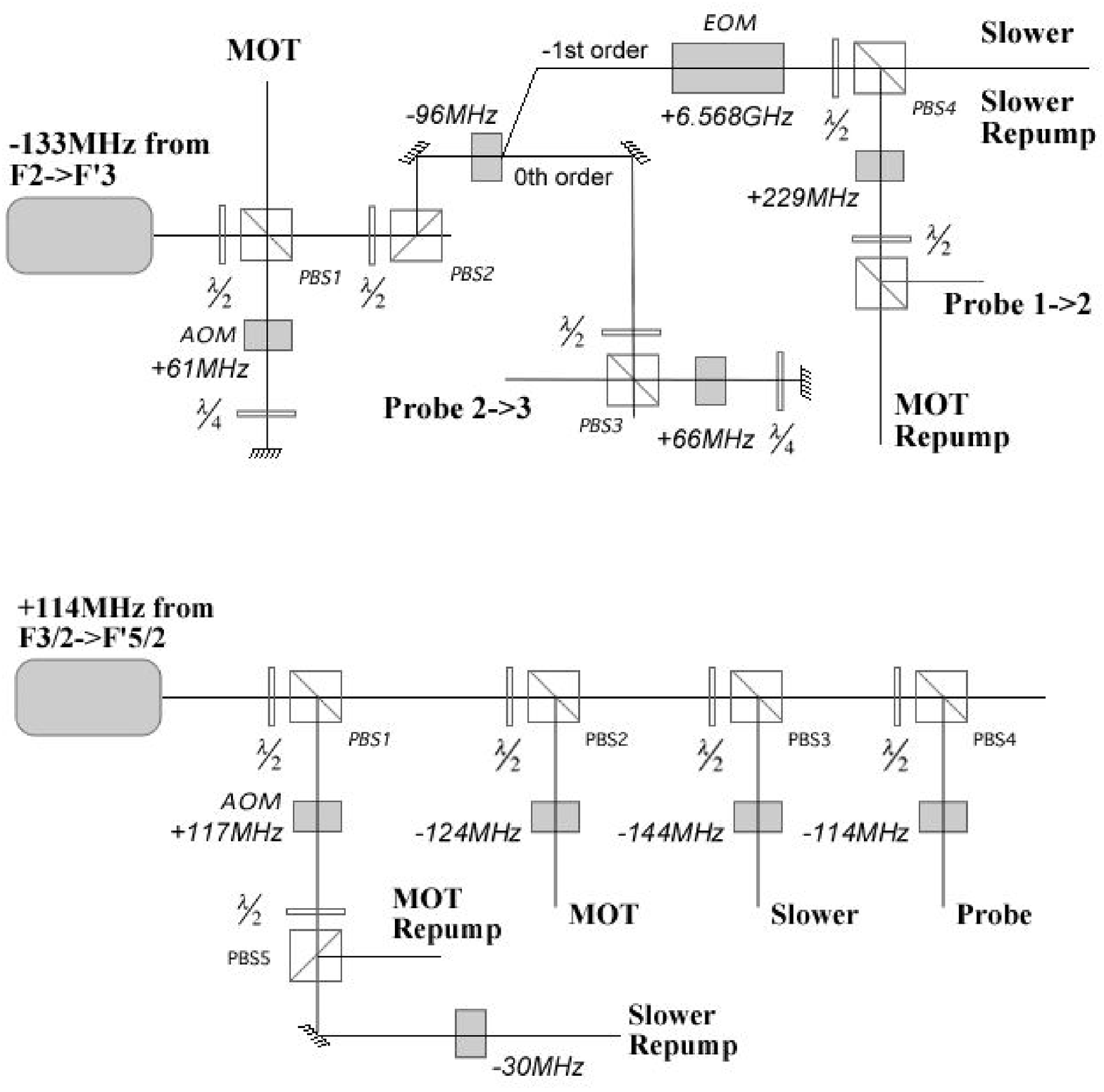}
\caption{Frequency synthesis schemes.  For both species, we use 20 mW for the Slower beam 
and 10 mW for each MOT beam, and in order to keep the transitions closed, repumping beams for 
the MOT and Slower with $5 \%$ power are also generated. For $^{87}$Rb (above), with AOMs 
and EOM providing the required detunings, the MOT and Slower beams are detuned by -11 MHz 
and -229 MHz from the F=2$\to$F$^\prime$=3 transition, respectively.  MOT and Slower 
repumping beams are in resonance with and detuned by -229 MHz from the F=1$\to$F$^\prime$=2 
transition. For $^{6}$Li (below), the MOT and Slower beams are detuned by -10 MHz and -30 MHz 
from the F=3/2$\to$F$^\prime$=5/2 transition; repumping beams for the MOT and Slower are 
in resonance with and detuned by -30 MHZ from the F=1/2$\to$F$^\prime$=3/2 transition. 
Probe beams are also generated which are in resonance with each of the selected transitions.}
\label{freqsynth}
\end{figure}

\subsection{Lasers and optics}
\label{optics}

For laser cooling of both species we use high-power diode lasers, 350 mW at a wavelength of 780 nm 
for  $^{87}$Rb and 400 mW at 671 nm for $^{6}$Li (see Fig. \ref{LaserPhoto} for an overview).  
For each laser the auxiliary beam (with power on the order of a few mW) is used for saturation 
absorption spectroscopy to achieve frequency locking, while the main beam supplies the Zeeman, 
MOT, probe, and repumping lines.  For $^{87}$Rb a commercial vapor cell at room temperature is 
used, while for $^{6}$Li we built a vapor cell with operating temperature 
$\sim 400^\circ$C \cite{li_vaporcell}.  All the necessary frequency detunings are realized 
by sending the main beams through various acousto-optic  modulators (AOMs) and, in the case 
of rubidium, also an electro-optic modulator (EOM), with single-pass and double-pass AOM 
efficiencies in the $65-75\%$ and $30-40\%$ ranges, respectively, and repumping EOM 
efficiency around $3\%$.  Once all the beams are properly detuned and polarized, they 
are delivered to the apparatus. Dichroic mirrors are used to combine the two sets of six 
beams necessary for simultaneous magneto-optical trapping of $^{6}$Li and $^{87}$Rb. 

For $^{87}$Rb, a Toptica DLX110 tunable high power diode laser is locked to the cross-over resonance 
between the F=2$\to$F$^\prime$=3 and F=2$\to$F$^\prime$=2 transitions at 780nm.  The laser setup for 
$^{6}$Li is similar to that for $^{87}$Rb, with a Toptica TA100 amplified tunable diode laser which 
is locked to the cross-over resonance between $^{6}$Li F=3/2$\to$ F$^\prime$=5/2 and 
F=1/2$\to$F$^\prime$=5/2 transitions at 671 nm.  Figure \ref{freqsynth} shows the frequency 
synthesis scheme for all required beams for both species, and the caption includes details 
of beam power budget and detunings.
 
\section{Design of magneto-optical and light-assisted magnetic traps} 

The Zeeman slowed $^{6}$Li and $^{87}$Rb atoms are captured, mixed and pre-cooled in the MOT. 
A pair of coils in anti-Helmholtz configuration produces an axial magnetic field gradient of 8 G/cm.  
The six-beam configuration is beneficial for achieving sub-Doppler optical molasses. 
After pre-cooling and hyperfine state preparation, we will transfer the atoms to a magnetic trap 
using $^{6}$Li in the F=3/2, $m_{F}$=3/2 state and either F=2, $m_{F}$=2 or F=1, $m_{F}$=-1 for 
$^{87}$Rb, all of which are weak-field seeking states.  The stretched states (as used in 
\cite{Silber}) have the advantage of minimizing spin-exchange losses while the 
$^{6}$Li (3/2,3/2)-$^{87}$Rb(1,-1) combination yields the maximum natural trapping 
frequency ratio, more than twice that of the  $^6$Li (3/2,3/2)-$^{23}$Na(2,2) 
combination that has yielded a $T/T_{\mathrm F}=5 \times 10^{-2}$ \cite{Hadzibabic}.  
Our trapping scheme is inspired by the design of an optically plugged quadrupole trap 
(OPT) \cite{plugged}, but is designed for a lower power deconfining laser beam.  
The coils form a quadrupole magnetic field which traps low-field seeking particles, but 
suffers from Majorana loss due to the magnetic field zero at the center. 
However, in the presence of a blue-detuned laser beam, the atoms are repelled from 
the center of the trap and so spin-flip related losses are reduced.  
The trapped rubidium will be then evaporatively cooled, in turn sympathetically cooling 
the fermionic lithium.  As discussed above, it is advantageous to prevent the rubidium 
from entering a deeply Bose condensed phase, as the gas has a much higher heat capacity 
in the classical regime and around the Bose-Einstein phase transition.

\begin{figure}[t]
\includegraphics[width=1.00\columnwidth]{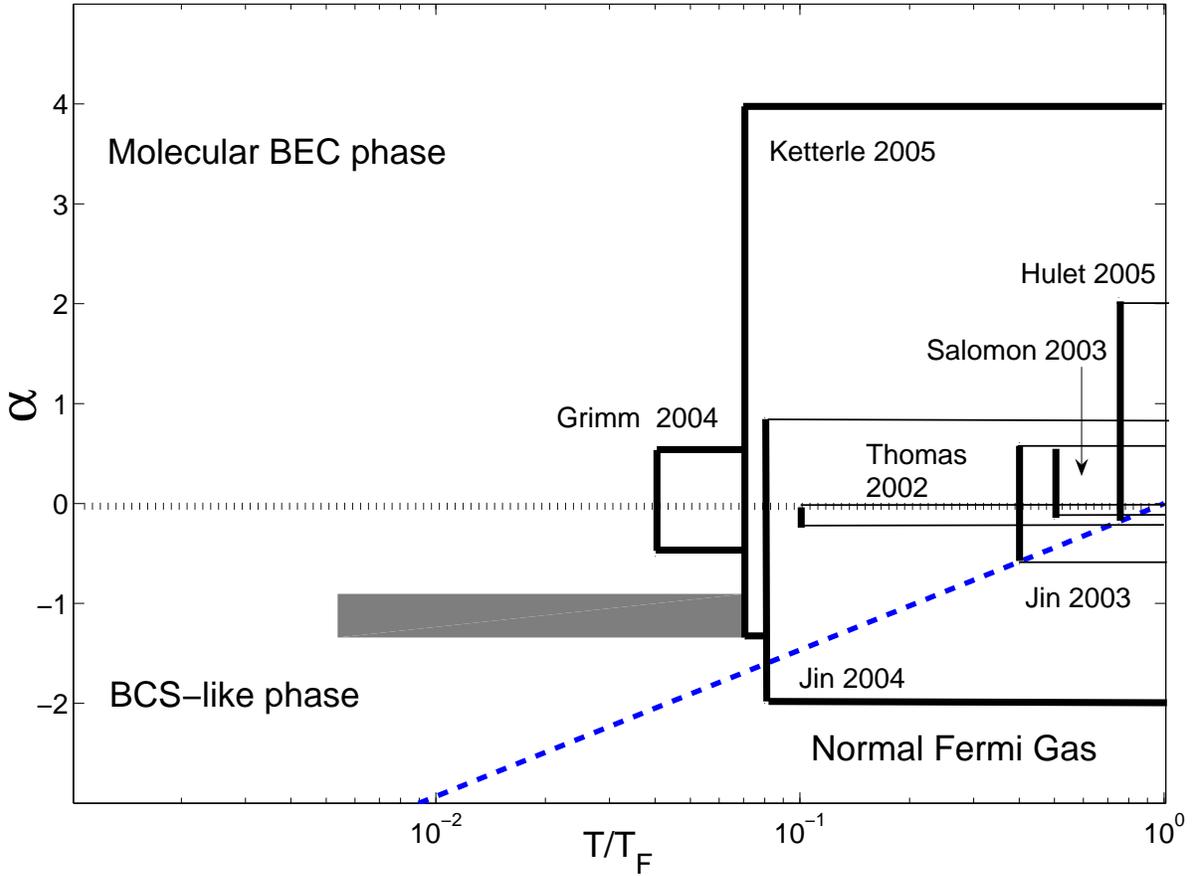}
\caption{Current status of Fermi degeneracy achieved by various groups and indication 
of our target domain (shaded area). The coupling strength parameter  $\alpha=1/\mathrm{k_{F}}a$, 
with $\mathrm{k_F}$ the Fermi wavenumber and $a$ the scattering length, is plotted versus 
the degeneracy parameter $T/T_{\mathrm F }$.  The dotted horizontal line at $\alpha=0$ 
separates the parameter space into the two regions of BEC molecular regime ($\alpha>0$) 
and BCS regime ($\alpha<0$). 
The dashed line, based on Eq. (3) in \cite{Houbiers}, provides further demarcation between the 
BCS superfluid phase (based on a critical temperature arising from s-wave pairing in the different 
internal states) and the normal Fermi gas phase. Our targeted region extends into 
$T/T_{\mathrm F }$=5$\times10^{-3}$, well within the Fermi degeneracy on the BCS side.  
The spread along the $\alpha$ direction should be obtained, rather than by changing 
$a$ with Feshbach resonances, by modifying $\mathrm{k_{F}}$ via different trapping strengths, 
varying the intensity of the deconfining laser beam.  Areas labeled Grimm 2004, Ketterle 2005, 
Jin 2004, Thomas 2002, Jin 2003, Salomon 2003, and Hulet 2005 are from \cite{Grimm}, \cite{Zwierlein1}, 
\cite{Jin1}, \cite{Thomas1, Thomas2}, \cite{Jin2}, \cite{Salomon}, and \cite{Hulet}, respectively.  
A degeneracy factor of $T/T_{\mathrm F }$=5$\times 10^{-2}$ was reported in \cite{Hadzibabic} 
without Feshbach modulation of the scattering length.}
\label{andyplot}
\end{figure}

In the OPT we are currently building, each coil has 6 layers of 10 windings of 1/8 inch square 
copper tubing. The inner diameter of the coils is 3 inches, and the spacing between the two coils 
is 2 inches. This pair of coils will be used to generate the magnetic gradient field for both MOT 
and OPT by quickly switching the current from 10 A to 200 A, which results in field gradients of 
20 G/cm and 400 G/cm respectively.  A 690 nm laser provides a 35 mW beam which propagates along the 
axis of the coils and is focused into the center of the quadrupole trap with a waist of 40 $\mu$m.  
Use of a laser beam with a wavelength in between the atomic transitions for the two species 
(red-detuned for $^{6}$Li and blue-detuned $^{87}$Rb) will result in stronger confinement of 
the fermionic species, at the price of increased localization of the lithium atoms near the trap 
center, where Majorana spin-flip losses become significant.  However, in the degenerate regime 
the Fermi gas will experience an effective Pauli repulsion and therefore a relatively smaller number 
of atoms around the trap center will be lost through Majorana spin-flips. 

\section{Conclusions}
The minimum reachable Fermi degeneracy in experiments using sympathetic cooling techniques 
is limited by the heat capacity matching of the two species. We have described an apparatus 
under development to explore these concepts and aimed at reaching deeper Fermi degeneracy.  
The achievement of lower $T/T_{\mathrm F }$ is desirable for the further exploration of the 
BEC-BCS crossover regime.  Specifically, we aim to broaden the experimental parameter space 
to study Fermi pairing into various superfluid phases.  The parameter space of $\alpha=1/\mathrm{k_{F}}a$ 
versus $T/T_{\mathrm F }$ achieved in various ongoing experiments is shown in Fig. \ref{andyplot}, 
also including the region we propose to study. The latter is complementary to the former, in  
the sense that we do not intend to explore a large region of the coupling parameter  
$\alpha$ at least as far as a magnetic trap is used. We should reach 
$T/T_{\mathrm F }$ ratios  an order of magnitude lower than those currently achieved.
New phenomena, including a non-trivial phase for superfluidity only present at 
finite temperature \cite{Chien,LevinPRA}, are expected in the $T/T_{\mathrm F }$ 
regime which could be accessible with our technique.  The use of a bichromatic 
optical trap configuration as discussed in \cite{Onofrio} could extend the study 
to a broader $\alpha$ range through Feschbach resonances.

As a concrete example of a research direction we can investigate the role played by 
the effective mass of fermions for superfluid pairing in optical lattices.  
Mismatched pairing has been suggested to occur as a result of unequal cloud 
densities or masses, for example a mixture of $^{6}$Li and $^{40}$K \cite{Wilczek}.  
This mass difference could also be realized by preparing the fermions in two 
separate hyperfine states which have significantly different electric dipole 
polarizabilities.  In the presence of an optical lattice this would correspond 
to the two states having different effective masses.  We have examined the case 
of two lithium hyperfine states ($2S_{1/2}^{2}$ and $2P_{1/2}^{2}$) and an effective 
mass ratio dependent on the two electric polarizabilities.  It appears that state-selective magnetic 
trapping is possible with $^6$Li hyperfine states and could allow for the 
comparison between various exotic states of superfluids, including LOFF states 
\cite{LOFF} and their combinations \cite{Dukelsky}.

More generally, superfluid Fermi-Bose mixtures may be used as an analog computer 
to solve equations mimicking the equations of state for quantum chromodynamics in regimes where 
perturbative techniques or lattice computations have not yet succeeded.
First theoretical steps in this very promising and interdisciplinary direction 
can be found in \cite{He1,Rapp,He2,He3}.

\begin{acknowledgments}
We are grateful to Richard L. Johnson for skillful technical support
and high-quality machining, and to David Collins for electronics
design and support. MBH, QW, and WJK acknowledge support from the Dartmouth 
Graduate Fellowship.  MBH has also been supported by the NSF-GAANN and Gordon 
Hull Fellowship programs.  RO acknowledges support from Dartmouth and from 
MIUR, Italy, under PRIN 2004028108$\_{001}$.
\end{acknowledgments}

\end{document}